\documentclass[10pt, conference, compsocconf]{IEEEtran}

\IEEEoverridecommandlockouts

\usepackage[ascii]{inputenc}
\usepackage[T1]{fontenc}
\usepackage{graphicx}  
\usepackage{amsmath,amssymb}
\usepackage{cite}
\usepackage{accsupp}
\usepackage{listings}
\usepackage{xcolor}
\usepackage[caption=false]{subfig}  
\usepackage{wrapfig}
\usepackage{url}
\newcommand{\lstbackground}{white}

\newcommand{\code}{\lstinline}

{}\newif\ifcomment  
\newcommand{\Cpp}{%
  C\kern-0.05em\protect\raisebox{.35ex}{$\scriptstyle+$\kern-0.05em$\scriptstyle+$}%
}

\begin{document}

\title{External Memory Pipelining Made Easy With TPIE}

\newcommand{\madalgomark}{$^1$}
\newcommand{\madalgothanks}{%
1: Supported in part by the Danish National Research Foundation
grant DNRF84
and Innovation Fond Denmark.
MADALGO is the Center for Massive Data Algorithmics,
a center of the Danish National Research Foundation.
}
\newcommand{\madalgoaddr}{%
  \affaddr{%
    MADALGO\\
    Aarhus University\\
    Aarhus, Denmark
  }
}
\newcommand{\scalgoaddr}{%
  \affaddr{%
    SCALGO\\
    Aarhus, Denmark
  }
}
\author{\IEEEauthorblockN{%
  Lars Arge\madalgomark{}%
    \thanks{%
\madalgothanks
    }%
  , %
  Mathias Rav\madalgomark{}%
  , %
  Svend C.\@ Svendsen%
  \madalgomark{}%
  }
  \IEEEauthorblockA{%
    MADALGO\\
    Aarhus University\\
    Aarhus, Denmark\\
    \{\texttt{large,rav,svendcs}\}\texttt{@madalgo.au.dk}
  }
  \and
  \IEEEauthorblockN{%
    Jakob Truelsen%
  }
  \IEEEauthorblockA{
    SCALGO\\
    Aarhus, Denmark\\
    \texttt{jakob@scalgo.com}
  }
}
\maketitle

\begin{abstract}
When handling large datasets that exceed the capacity of the main memory,
movement of data between main memory and external memory (disk), rather
than actual (CPU) computation time, is often the bottleneck in the
computation. Since data is moved between disk and main memory in large
contiguous blocks, this has led to the development of a large number of
I/O-efficient algorithms that minimize the number of such
block movements. However, actually implementing these algorithms can be
somewhat of a challenge since operating systems do not give complete
control over movement of blocks and management of main memory.

TPIE is one of two major libraries that have been developed to support
I/O-efficient algorithm implementations. It relies heavily on the fact
that most I/O-efficient algorithms are naturally composed of components
that stream through one or more lists of data items while producing one
or more such output lists, or components that sort such lists. Thus
TPIE provides an interface where list stream processing and sorting can
be implemented in a simple and modular way without having to worry about
memory management or block movement. However, if care is not taken, such
streaming-based implementations can lead to practically
inefficient algorithms since lists of data items are typically written to
(and read from) disk between components.

In this paper we present a major extension of the TPIE library that
includes a pipelining framework that allows for practically efficient
streaming-based implementations while minimizing I/O-overhead between
streaming components. The framework pipelines streaming components to
avoid I/Os between components, that is, it processes several components
simultaneously while passing output from one component directly to the
input of the next component in main memory. TPIE automatically
determines which components to pipeline and performs
the required main memory management, and the extension also includes
support for parallelization of internal memory computation and progress
tracking across an entire application. Thus TPIE supports efficient
streaming-based implementations of I/O-efficient algorithms in a simple,
modular and maintainable way. The extended library has already been used
to evaluate I/O-efficient algorithms in the research literature and is
heavily used in I/O-efficient commercial terrain processing applications
by the Danish startup SCALGO.
\end{abstract}

\section{Introduction}
\label{sec:introduction}

When handling large datasets that exceed the capacity of the main memory,
movement of data between main memory and external memory (disk), rather than
actual (CPU) computation time, is often the bottleneck in the computation.
The reason for this is that disk access is orders of magnitude slower than
internal memory access. Thus, since data is moved between disk and main memory
in large contiguous blocks, it is often more important to design algorithms
that minimize block movement than computation time when handling massive data.
This has led to the development of a large number of
\emph{I/O-efficient algorithms} in the I/O-model by Aggarwal and Vitter \cite{AV88}.
In this model, the computer is equipped with a two-level memory hierarchy
consisting of an internal memory capable of holding $M$ data items and an
external memory of conceptually unlimited size. All computation has to happen
on data in internal memory and data is transferred between internal and
external memory in blocks of $B$ consecutive data items. Such a transfer is
called an \emph{I/O-operation} or \emph{I/O}, and the cost of an algorithm is
the number of I/Os it performs.
The number of I/Os required to read or write $N$ items from disk is
$\mathrm{Scan}(N) = \lceil N/B\rceil$, while the number of I/Os required to sort
$N$ items is $\Theta(\mathrm{Sort}(N)) = \Theta((N/B) \log_{M/B}(N/B))$~\cite{AV88}.

While many I/O-efficient algorithms have been developed in the I/O-model of
computation, actually implementing these algorithms can be somewhat of a
challenge since operating systems do not give complete control over movement of
blocks and management of main memory.
However, two major libraries TPIE \cite{tpie12} and STXXL \cite{DKS08} have
been developed to support I/O-efficient algorithm implementations.
It turns out that most I/O-efficient algorithms are naturally composed of
components that stream through one or more lists of data items while producing
one or more such output lists, or components that sort such lists.
TPIE in particular uses this to provide an interface where list stream
processing and sorting can be implemented in a simple and modular way without
having to worry about memory management or block movement.
However, if care is not taken, such a streaming-based implementation can lead
to practically inefficient algorithms since lists of data items are typically
written to (and read from) disk between components.
In implementations consisting of many small (but I/O-efficient) components, the
I/Os incurred when writing and reading such lists can easily comprise more than
half of the total number of I/Os, and while this may not be a problem when
considering asymptotic theoretical performance, it is unacceptable in practice
when the total execution time is measured in hours or days.

In this paper we present a major extension of the TPIE library that includes a
pipelining framework that allows for practically efficient streaming-based
implementations while minimizing I/O-overhead between streaming components.
The framework pipelines streaming components to avoid I/Os between components,
that is, it processes several components simultaneously while passing output
from one component directly to the input of the next component in main memory.
TPIE automatically determines which components to pipeline and performs the
required main memory management, and the extension also includes support for
parallelization of internal memory computation and progress tracking across an
entire application. Thus TPIE supports efficient streaming-based
implementations of I/O-efficient algorithms, and TPIE applications are
naturally implemented as reusable components, thereby reducing programming time
and code duplication.

\subsection{Previous Work}
\label{sec:previous-work}

As mentioned, two major software libraries support implementation of
I/O-efficient algorithms for big data analysis, namely TPIE \cite{tpie12} and STXXL \cite{DKS08}.
They are both {\Cpp} software libraries,
and as opposed to many of the frameworks that have emerged for supporting big data analysis in
the last decade, such as e.g.\@ MapReduce~\cite{mapreduce08}, Spark~\cite{spark10}, and Flink~\cite{flink14},
they mainly support single-host implementations.
One reason for this is that the libraries, in particular TPIE,
are designed to support implementations on standard commodity hardware.
Another reason is that no efficient distributed algorithms are known for many of the problems for
which I/O-efficient algorithms have been studied and implemented;
we refer to surveys \cite{argesurvey,vittersurvey} and descriptions of implementations
(e.g.\@ \cite{ADMO2006,ABTT2013,ATY2014,DSSS2004,MO2009}) for references.
Thus in this paper we also focus on single-host implementations.
However, is should be mentioned that in the context of distributed programming,
pipelining has recently been studied with the Thrill framework~\cite{thrill16}.

Although both libraries for implementation of I/O-efficient algorithms,
the overall philosophies of the TPIE and STXXL are somewhat different.
The philosophy of TPIE (the Templated Portable I/O Environment) is to provide a high-level interface that allows for easy translation of abstract I/O-efficient algorithm descriptions into code that is portable across computational platforms and not unnecessarily complex. Thus building on the fact that most I/O-efficient algorithms are composed of streaming components, TPIE provides a generic stream interface that hides how blocked I/O is performed and instead provides methods for processing one data item at a time. TPIE also provides internal memory management where memory allocations are automatically counted towards an application-wide memory limit, and where an application can at any point determine the currently available main memory. Thus applications do not have to explicitly keep track of available memory, which often simplifies implementations considerably. For example, in the TPIE built-in streaming-based implementation of the I/O-optimal $\mathcal{O}(\mathrm{Sort}(N))$ external multi-way merge-sort, the number of sorted streams that can be merged I/O-efficiently (without being swapped out by the operating system) depends on the available main memory, where care has to be taken to ensure that the memory used to hold blocks of items for each used stream is counted towards the amount of available memory. The TPIE memory management allows for determining the number of streams to merge without explicitly keeping track of available memory and memory used for blocked I/O. Overall, TPIE is designed to remove focus from the tedious details of creating I/O-efficient applications and allows for implementations that are efficient on all hardware platforms with minimal configuration.

The philosophy of STXXL (Standard Template library for XXL data sets) on the other hand is to achieve maximum I/O-throughput by reducing I/O-overhead as much as possible and by exposing the characteristics of the hardware to the application programmer. Thus, to avoid any overhead induced by the operating system, STXXL allows the user to configure separate disks for use with applications outside of the file systems of the operating system. In fact, STXXL project programmers recommend that a separate disk is set aside for STXXL applications. STXXL also explicitly supports parallel disks. Like TPIE, STXXL supports streaming-based implementations and includes various basic streaming components such as sorting, but unlike TPIE it actually contains support for pipelining of streaming components. However, STXXL expects the application programmer to explicitly define which components to pipeline and explicitly manage main memory. Thus, the programmer e.g.\@ has to specify how much memory each streaming component in a pipelined application should use. A separate (not officially released) branch of STXXL contains support for utilizing multi-core processors for the internal-memory work of pipelined applications \cite{BDS2009}.
Overall, STXXL is designed such that an application can be tailored to the available hardware, and with the proper configuration an STXXL application can
achieve close to full utilization of the available I/O bandwidth.

\subsection{Our Results}
\label{sec:results}

In this paper we present a major extension of the TPIE library that includes a pipelining framework that allows for practically efficient streaming-based implementations while minimizing I/O-overhead between streaming components. The extension also includes support for progress tracking across an entire application and for parallelization of internal memory computation.

Like STXXL, the TPIE pipelining framework saves I/Os by passing intermediate results between streaming components directly in main memory. However, TPIE pipelining is the first framework to provide automatic pipeline and memory management and thus combining the best of the TPIE and STXXL streaming philosophies. The framework is component-centric in that the memory requirement of each streaming component is specified locally by the component developer. The automatic pipeline and memory management then means that at runtime TPIE will automatically determine which components to pipeline and distribute memory among multiple components of a large application in a way that automatically uses all the main memory available to the application.
Thus, unlike in STXXL, a TPIE programmer e.g.\@ does not have to consider how the memory uses of the individual components have to be adjusted when they are combined into an application. While such adjustments, along with adjustments of the grouping of components into pipelines, can be done manually for small projects, it can be very cumbersome for large-scale professional software projects involving many programmers, where modification of a component to use more memory can very easily lead to memory over-usage problems if the memory use of other components are not adjusted accordingly. Thus the TPIE component-centric approach simplifies the application development process, promotes modularity and supports maintainability.

Since I/O-efficient applications are typically long-running processes that take hours or days to complete, it is important that an application is able to provide a progress bar that gives a precise estimate of the progress of its execution. To be able to do so in a simple way, the TPIE extension also takes a component-centric approach. Like for memory use, a component developer can in a simple way include support for information about the progress of the component, and TPIE automatically combines information from all components and thus supports a single progress bar that advances from 0\% to 100\% at a constant pace. Thus, again the use of a component-centric approach promotes modularity.

Especially after minimizing I/O, the use of multi-core parallelization can often help to bring down the running time of massive data applications. Thus the TPIE extension includes easy support for such parallelization by allowing the application programmer to wrap a part of a pipeline in a parallelization directive that will
trivially parallelize it across all CPU cores.
For instance, when forming sorted runs in multi-way merge sort, the internal memory sorting algorithm in TPIE automatically uses all the available CPU cores.

Overall, the major TPIE library extension presented in this paper supports efficient streaming-based implementations of I/O-efficient algorithm in a simple, modular and maintainable way, and I/O-efficient algorithms can thus be composed and adapted in commercial and research applications while dealing systematically with important aspects, such as memory management and progress tracking, that are not intrinsic to the algorithmically optimal solution. The extended library has already been used to evaluate I/O-efficient algorithms in the research literature (e.g.\@ \cite{ABTT2013,ATY2014}) and is heavily used in I/O-efficient commercial terrain processing applications by the Danish startup SCALGO\footnote{%
  SCALGO: Scalable Algorithmics. \url{https://scalgo.com}}.
The extension is integrated into the official TPIE project that is available on GitHub as free and open source software\footnote{%
  TPIE: Templated Portable I/O Environment. \url{http://madalgo.au.dk/tpie}}.

The rest of the paper is structured as follows.
In Section~\ref{sec:example}, we motivate pipelining with a concrete algorithm
based on scanning and sorting.
After this motivation, we in Section~\ref{sec:concepts} present in full generality
how to use the TPIE extension and briefly discuss its implementation.
\begin{figure}
  \centering%
  \hspace*{-0.7cm}%
  \includegraphics[width=\dimexpr\linewidth+1.4cm\relax]{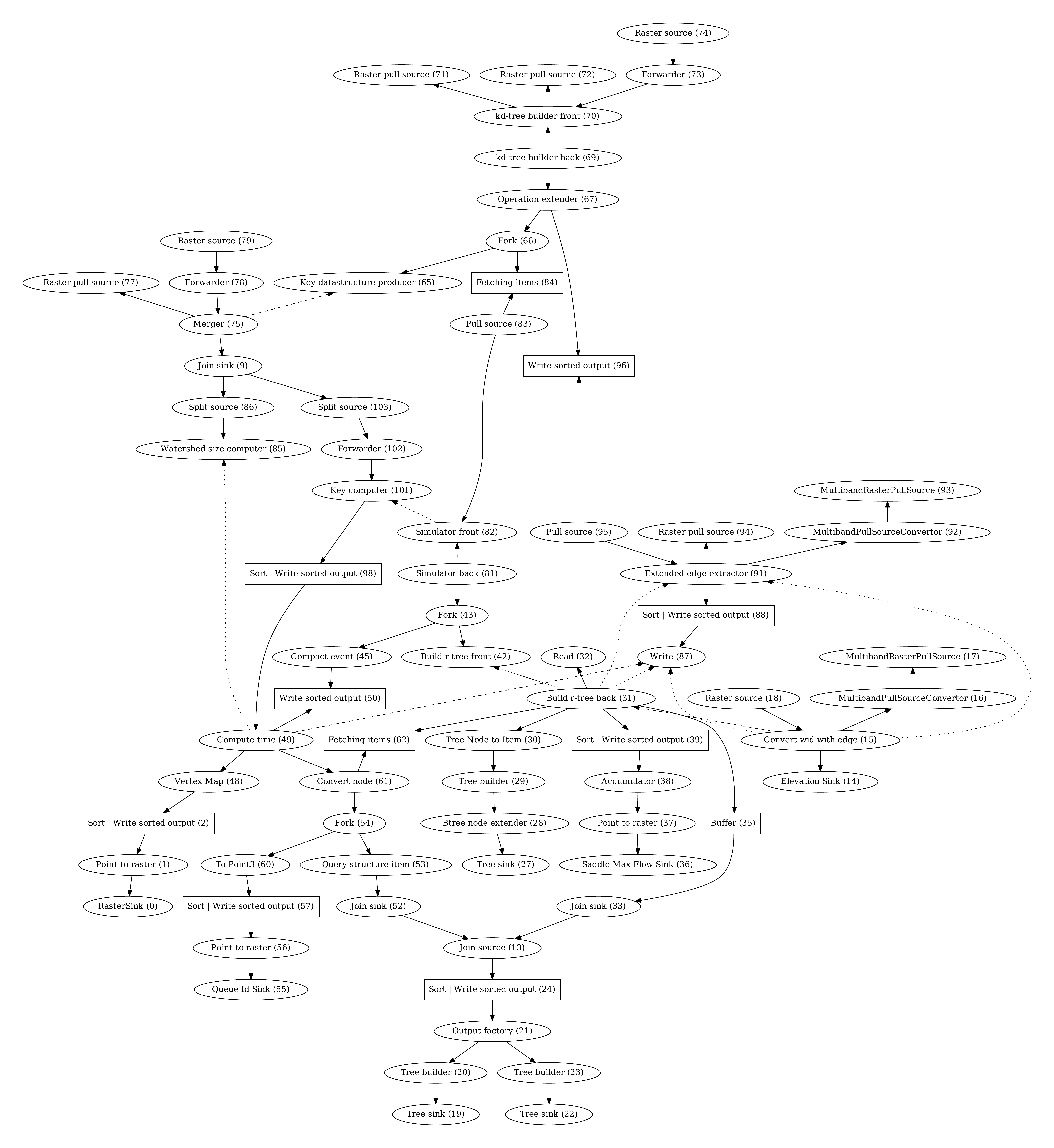}%
  \hspace*{-0.7cm}%
  \caption{
    Pipelined components in a real TPIE application developed by SCALGO.
    Dashed lines represent phase ordering dependencies resulting from blocking edges in the flow graph.
  }
  \label{fig:scalgo-pipe}
\end{figure}

\section{An Example Problem}
\label{sec:example}

In this section we present an example of a typical sub-problem in an I/O-efficient application,
and we show that the problem benefits from a pipelined implementation.
By implementing every sub-problem in a bigger data processing application
(such as the real-world example in Figure~\ref{fig:scalgo-pipe})
using pipelining, more than half of the I/Os overall can be saved.

\subsection{The Raster Transformation Problem}
\label{sec:example-def}

In geographic information systems (GIS), a terrain is often represented as a raster of heights with each cell indicating the height of the terrain in a certain point. Since the Earth is spherical and a raster is flat, it is not possible to map the entire surface of the Earth continuously to a raster. However, if only a particular region, country or continent needs to be represented, it is possible to project the chosen region to a plane in a way that roughly maintains the geodesic distances and areas. When several rasters must be processed together they must be in the same projection.

We call the problem of transforming a raster from one projection to another the \emph{raster transformation problem}. Essentially, the problem consists of projecting each cell of the raster from the source projection plane to the unit sphere and from the unit sphere to the target projection plane. These two steps can be represented by a function $f : \mathbb{Z}^2 \to \mathbb{Z}^2$ that maps each cell of the \emph{target} raster projection to the corresponding cell of the \emph{source} raster projection. Thus, in the raster transformation problem we are given an input raster $A$ of size $W \times H$ (that is a $W$ by $H$ matrix of numbers) stored in row-major order, and we want to produce an output raster $B$ of size $W' \times H'$ in row-major order such that the value of a cell $(x,y)$ in $B$ is copied from the value of a cell $(x',y') = f(x,y)$ in $A$. Below we for convenience let $N = W H = W' H'$ be the number of cells in both the input and output raster.

\begin{figure}[bt]
  \begin{minipage}{0.5\linewidth}%
    \leavevmode
    \subfloat[]{%
      \includegraphics[page=1]{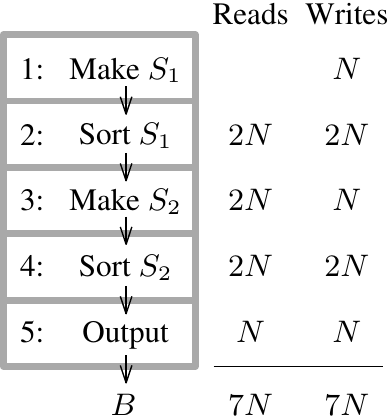}%
      \label{fig:algorithm-old}%
    }%
    \hspace*{\fill}%
  \end{minipage}%
  \hspace*{\fill}%
  \begin{minipage}{0.5\linewidth}%
    \leavevmode
    \hspace*{\fill}%
    \subfloat[]{%
      \includegraphics[page=2]{example-onecol.pdf}%
      \label{fig:algorithm-new}%
    }%
  \end{minipage}%
  \\%
  \caption{%
    Raster transformation algorithm.
    \protect\subref{fig:algorithm-old}~%
    The algorithm without pipelining, requiring $7N$ reads and writes.
    \protect\subref{fig:algorithm-new}~%
    The algorithm with pipelining, requiring just $3N$ reads and writes.
  }
  \label{fig:algorithm}
\end{figure}

The raster transformation problem can easily be solved in optimal $\mathcal{O}(N)$ time simply by for each cell $(x, y)$ in $B$ reading the corresponding input value at $f(x, y)$ in $A$. However, this solution might be very I/O-inefficient. For example, if $f$ represents matrix transposition where $f(x, y) = (y, x)$ then each access to $A$ requires a new block to be read (assuming $W, H \geq \frac{M}{B}$) and thus the solution performs $\Theta(N)$ I/Os in the worst case. For matrix transposition in particular, only $\Theta(\mathrm{Sort}(N))$ I/Os are required \cite{AV88}. In fact, in general the problem can be solved in $\mathcal{O}(\mathrm{Sort}(N))$ I/Os using a simple five step streaming algorithm (refer to Figure~\ref{fig:algorithm-old}): First a stream $S_1$ is constructed containing for each cell $(x, y)$ of $B$ an item consisting of a pair $(f(x, y), (x, y))$. Next $S_1$ is sorted such that $f(x, y)$ appear in the same row-major order used to store $A$. In the third step, $A$ and $S_1$ are then scanned simultaneously to construct a stream $S_2$ containing an item $(x, y, v)$ for each pair $(f(x, y), (x, y))$ in $S_1$ where $v$ is the value of $A$ at position $f(x, y)$. Then $S_2$ is sorted into the row-major order used to store $B$. In the fifth and final step, $S_2$ is then scanned and for each entry $(x, y, v)$ the value $v$ is output to $B[x, y]$. Since the algorithm performs a constant number of scanning and sorting steps it uses $\mathcal{O}(\mathrm{Sort}(N))$ I/Os and can easily be implemented using the streaming support of either TPIE or STXXL.
Refer to GitHub\footnote{\url{https://github.com/Mortal/pipelining/blob/8996e5c87d/tpie_imperative/transform_paper.cpp#L60-L104}} for a TPIE code example.

As discussed in the introduction, streaming-based implementations of even simple I/O-efficient algorithms, as the raster transformation algorithm above, might not be practically efficient because items are written to disk between steps. To illustrate this, we will analyze the exact number of items read and written by the above algorithm. For simplicity, we assume that $N$ elements can be sorted using $2N$ reads and $2N$ writes, which is a practically realistic assumption if external merge-sort is used. Recall that external merge-sort works by first scanning through the $N$ input elements and sorting $M$ elements at a time in internal memory to produce $\frac{N}{M}$ sorted runs. This requires $N$ reads and $N$ writes. Next the sorted runs are merged together $\frac{M}{B}$ at a time (using a block of internal memory for each run) to produce $\frac{N}{M}/\frac{M}{B}$ longer sorted runs, again using $N$ reads and $N$ writes. The merging process is repeated until a single sorted output is obtained. However, in practice (where $N$,~$M$ and $\frac{M}{B}$ are on the order of $10^{12}$, $10^9$ and $10^3$, respectively) $\frac{N}{M}/\frac{M}{B}<1$ so only a single merging step is required. Using this, we can easily realize that the above raster transformation algorithm requires $7 N$ reads and $7 N$ writes without pipelining (refer again to Figure~\ref{fig:algorithm-old}): Generating the stream $S_1$ in the first step requires $N$ writes, and sorting it in the second step requires $2 N$ reads and $2 N$ writes. Reading $A$ and $S_1$ simultaneously in the third step to produce $S_2$ requires $2N$ reads and $N$ writes. Again, sorting $S_2$ in the fourth step requires $2N$ reads and writes, and finally, reading $S_2$ and writing $B$ in the fifth step requires $N$ reads and writes.
However, by modifying the five steps of the algorithm so that the intermediate result of one step is immediately used by the next step (if possible) without storing the intermediate result on disk, that is, by using pipelining, we can reduce the number of reads and writes to $3N$ each. More precisely, we can save $N$ writes and $N$ reads of $S_1$ between step one and two by immediately producing the initial sorted runs of step two while performing step one. Similarly, we can save the $N$ writes and $N$ reads of $S_1$ between step two and three by performing step three (scanning $S_1$ and $A$) simultaneously with merging the sorted runs. Note that apart from the write and read between the run formation and merging in step two, we in this way avoid writing $S_1$ altogether. In a similar way, we can avoid writing $S_2$ and save $N$ reads and $N$ writes by also producing the initial sorted run of step four while performing step 3, as well as $N$ reads and $N$ writes by performing step five simultaneously with merging of the sorted runs in step 4. Altogether, we save $4N$ reads and $4N$ writes, that is, over half of the I/Os. Although this does not change the asymptotic I/O-complexity of the algorithm, it translates into a running time reduction of 22 hours if we assume an input size $N$ of 1~TB and a disk read/write speed of 100~MB/s.

Note that the pipelining process described above conceptually transforms the five-step algorithm into a three-phase algorithm as indicated in Figure~\ref{fig:algorithm-new}, where e.g.\@ the second phase consists of the merging part of the step two sorting, step three, and the run formation part of the step four sorting. One could of course implement the algorithm by implementing these three phases directly, that is, by implementing several special versions of external merge-sort (or rather, special run formation, merging, and merging-run formation implementations). However, this would not only be cumbersome, but also unacceptable from a software engineering point of view. Instead, direct support of pipelining where the five-step algorithm is automatically pipelined would be desirable. However, such a pipelining would require system support for identification of phases and careful memory management. For example, the merge and run formation parts of phase two of the three-phase algorithm would normally both require all of the main memory, so the memory somehow needs to be divided between the two parts. As described below, this is handled somewhat differently in STXXL and the new TPIE extension.
\begin{figure*}
\begin{lstlisting}
void transform(raster_input & input, raster_output & output, size_t memory_available) {
    /*\label{line:sxs-mem-begin}
    */// In phase 1, the single sorter can use all the available memory.
    const size_t phase1_sort_memory = memory_available;
    // In phase 2, the two sorters receive each half the available memory,
    // excluding the memory used to store a single block from the input.
    const size_t phase2_sort_memory = (memory_available - input.buffer_size()) / 2;
    // In phase 3, there is a single sorter and an output buffer.
    const size_t phase3_sort_memory = memory_available - output.buffer_size();/*
    \label{line:sxs-mem-end}*/
    /*\label{line:sxs-defs-begin}
    */GenerateOutputPoints output_points(output.dimensions());
    typedef TransformPoints<GenerateOutputPoints> TransformOutputPoints;
    TransformOutputPoints point_pairs(std::move(output_points), input.dimensions());
    typedef stxxl::stream::runs_creator<TransformOutputPoints, input_yorder> rc_type;
    rc_type rc(std::move(point_pairs), input_yorder(), phase1_sort_memory);
    typedef stxxl::stream::runs_merger<rc_type::sorted_runs_type, input_yorder> rm_type;
    // The following call to rc.result() executes the first phase.
    rm_type rm(rc.result(), input_yorder(), phase2_sort_memory);/*\label{line:sxs-result1}*/
    RasterReader input_raster_reader(input);
    typedef PointFiller<rm_type, RasterReader> FillOutputPoints;/*\label{line:sxs-filler-typedef}*/
    FillOutputPoints filler(std::move(rm), std::move(input_raster_reader));
    typedef stxxl::stream::runs_creator<FillOutputPoints, point::yorder> rc2_type;
    rc2_type rc2(std::move(filler), point::yorder(), phase2_sort_memory);
    typedef stxxl::stream::runs_merger<rc2_type::sorted_runs_type, point::yorder> rm2_type;
    // The following call to rc2.result() executes the second phase.
    rm2_type rm2(rc2.result(), point::yorder(), phase3_sort_memory);/*\label{line:sxs-result2}*/
    // The following call to write_raster() executes the third phase.
    write_raster(std::move(rm2), output);/*\label{line:sxs-defs-end}\label{line:sxs-exec}*/
}
\end{lstlisting}
\caption{Raster transformation using the STXXL streaming layer.}
\label{fig:stxxl-code}
\end{figure*}

\subsection{STXXL Implementation}
\label{sec:example-stxxl}

When implementing the raster transformation algorithm with pipelining using the STXXL streaming layer, the five steps of the algorithm are implemented individually as is natural from a software engineering point of view; we call each such individual part of a pipeline a \emph{component}. However, since STXXL does not handle memory management, the implementation that combines the components then has to identify the three phases of the algorithm explicitly and compute how much memory is allocated to each of the components of a phase.
Refer to Figure~\ref{fig:stxxl-code}
for STXXL code that implements this, that is, the main code that implements the five step algorithm in three phases (excluding the code for the individual components). The code illustrates how three phases are explicitly identified and memory allocated. For example, in phase two the memory available for the two sorting components (merging of step two and run formation of step four) is computed by setting aside a buffer of size $B$ of the available main memory for reading the input, and then share the remaining memory between the two sorting components. Concretely, the computation is performed with the statement:
\code{sort_memory = (memory_available - block_size) / 2}.
While identifying phases and allocating memory in this way is easy in our simple example algorithm, it is more difficult in larger applications such as the example given in
Figure~\ref{fig:scalgo-pipe}%
. Especially if more than one programmer is working on the application, it is difficult and error-prone to distribute memory correctly.

Apart from the complexity that the need for phase identification and memory allocation adds to pipelined STXXL code, there are also some {\Cpp} syntax issues that add to the code complexity. More precisely, the {\Cpp} syntax used is quite verbose since for technical reasons names of the components often need to be repeated. The reason is that STXXL combines pipelining components using a {\Cpp} feature known as \emph{template instantiation} that allows for the compiler to inline function calls between different components of the pipeline. For performance reasons, this is necessary when many small components are pipelined. However, the template instantiation syntax is not well-suited for use in large pipelined applications. As an example, consider the {\Cpp} statement
  \code{typedef TransformPoints<GenerateOutputPoints> Transform}\-\code{OutputPoints;}
 in the STXXL implementation of the raster transformation algorithm.
In this statement, the {\Cpp} language \code{typedef} statement is used to declare \code{Trans}\-\code{form}\-\code{Output}\-\code{Points} to be a \emph{type alias} for the \code{Trans}\-\code{form}\-\code{Points} component instantiated with the \code{Gene}\-\code{rate}\-\code{Out}\-\code{put}\-\code{Points} component. Such a type alias is needed for each component of the pipeline, and only when all the type aliases have been defined the individual component objects can be declared and constructed. In this way, the pipeline has to be defined both in terms of type aliases nested within each other and as actual component objects combined together.

\subsection{TPIE Implementation Using Pipelining}
\label{sec:example-tpie}

\newcommand{\tplineshelp}[1]{\ref{line:tp-#1-begin}-\ref{line:tp-#1-end}}
\newcommand{\tplines}{lines~\tplineshelp}
\newcommand{\tpline}[1]{line~\ref{line:tp-#1}}

\newcommand{\nodelineshelp}[1]{\ref{line:node-#1-begin}-\ref{line:node-#1-end}}
\newcommand{\nodelines}{lines~\nodelineshelp}
\newcommand{\nodeline}[1]{line~\ref{line:node-#1}}

As in the case of pipelined STXXL, in the implementation of the five step raster transformation algorithm using the extended TPIE library with pipelining, the components of the pipeline are implemented individually. However unlike in the STXXL implementation, the combination of the components becomes very simple since TPIE is component-centric and automatically identifies phases and performs memory management. To illustrate this, a diagram showing the eight components used to implement the five steps is given in Figure~\ref{fig:project-pipeline} along with the pipelining code in Figure~\ref{fig:project-code}. Note how the code in Figure~\ref{fig:project-code} \tplines{def} naturally corresponds to the pipeline in Figure~\ref{fig:project-pipeline}.
Note that the reading and writing of rasters are handled by two special components to separate the handling of specific raster formats from the algorithm, and how the two sorting components are implemented using two different built-in TPIE sorting components defined in lines~\ref{line:tp-sort1} and \ref{line:tp-sort2} on Figure~\ref{fig:project-code}. The reason two different sorter implementations are used (and that the pipeline is defined in two statements defining \code{p1} and \code{p2}, respectively) is that the output from the component sorting $S_1$ has to be read by the component constructing $S_2$ simultaneously with the output from the component reading the input raster $A$. Thus, the component constructing $S_2$ has to control when data is received from the sorting component which is done through \emph{pull-based streaming}. This functionality is implemented with a TPIE so-called \emph{passive sorter} with an input and an output part defined in \tpline{sort1}. On the other hand, the component sorting $S_2$ is a more traditional pipelined component that uses \emph{push-based streaming}, where input data is received from preceding component (in this case the component constructing $S_2$) when ready, and output data in turn pushed to the subsequent component. It is defined with an ordinary TPIE sorter in \tpline{sort2}.
As an example of a component implementation, the code implementing the component \code{generate_output_points} is given in Figure~\ref{fig:generate-output-points}.
The component contains a method \code{propagate()} that is called by TPIE when setting up the pipeline, and a method \code{go()} that is called by TPIE to execute the actual component.
The component also uses push-based streaming, and it pushes each produced element to the next component by calling the \code{push()} method of that component.
The details of how the push and pull mechanisms work will be discussed in Section~\ref{sec:concepts}, where the full TPIE pipelining framework and its implementation is described. Below we highlight some of the other framework features that are used in the raster transformation example.

\newsavebox{\projectcode}
\begin{lrbox}{\projectcode}%
\begin{lstlisting}
void transform(raster_input & A, raster_output & B,
/*\hphantom{%
void transform(}*/tpie::progress_indicator_base & pi) {
    /*\label{line:tp-def-begin}
    */auto sort_S1 = tpie::pipelining::passive_sorter<projected_point>();/*\label{line:tp-sort1}*/
    auto sort_S2 = tpie::pipelining::sort(point::yorder());/*\label{line:tp-sort2}*/
    tpie::pipelining::pipeline p1 = generate_output_points()/*\label{line:tp-generate}*/
        | tpie::pipelining::parallel(compute_transformation())/*\label{line:tp-parallel}*/
        | sort_S1.input();
    tpie::pipelining::pipeline p2 = read_raster(A)
        | construct_S2(sort_S1.output())
        | sort_S2/*\label{line:tp-push-sort}*/
        | construct_output()
        | write_raster(B);/*
    \label{line:tp-def-end}*/
    /*\label{line:tp-exec-begin}\label{line:tp-forward-begin}
    */p1.forward("inputsize", A.dimensions());
    p1.forward("outputsize", B.dimensions());/*
    \label{line:tp-forward-outputsize}\label{line:tp-forward-end}*/
    uint64_t n = A.cell_count() + B.cell_count();/*\label{line:tp-n}*/
    p1(n, pi, TPIE_FSI);  // Execute the pipeline/*\label{line:tp-exec}\label{line:tp-exec-end}*/
}
\end{lstlisting}%
\end{lrbox}
\begin{figure*}[bt]
  \begin{minipage}{200pt}%
    \subfloat[]{%
      \includegraphics{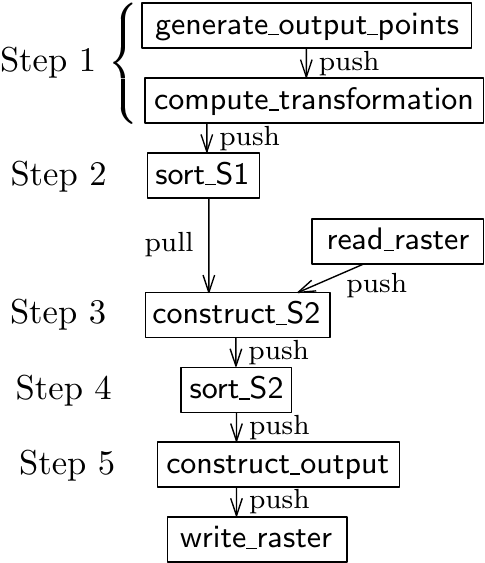}%
      \label{fig:project-pipeline}%
    }%
  \end{minipage}%
  \hspace*{\fill}%
  \begin{minipage}{277pt}%
    \subfloat[]{%
      \usebox{\projectcode}%
      \label{fig:project-code}%
    }%
  \end{minipage}%
  \\%
  \caption{%
    Raster transformation algorithm.
    \protect\subref{fig:project-pipeline}~%
    Pipeline illustrated as components.
    \protect\subref{fig:project-code}~%
    TPIE code implementing the pipeline.%
  }
  \label{fig:project}
\end{figure*}

\newcommand{\pipefeature}[1]{\textbf{#1.}}

\begin{figure}[t]
  \begin{lstlisting}[emph={fetch,set,step,steps,push,go,propagate},emphstyle={\bfseries}]
template <typename dest_t>
struct GenerateOutputPoints : public tpie::pipelining::node {
    GenerateOutputPoints(dest_t d): dest(std::move(d)) {}
    virtual void propagate() override {/*\label{line:node-propagate-begin}*/
        dimensions = fetch<rastersize_t>("outputsize");/*\label{line:node-fetch}*/
        set_steps(dimensions.width * dimensions.height);/*\label{line:node-set-steps}*/
    }/*\label{line:node-propagate-end}*/
    virtual void go() override {/*\label{line:node-go-begin}*/
        for (int y = 0; y < dimensions.height; y++)
            for (int x = 0; x < dimensions.width; x++)
                { step(); dest.push(point(x, y)); }/*\label{line:node-step}\label{line:node-push}*/
    }/*\label{line:node-go-end}*/
    rastersize_t dimensions; dest_t dest;
};
typedef tp::pipe_begin<tp::factory<GenerateOutputPoints>>
    generate_output_points;
\end{lstlisting}
  \caption{%
    \code{generate_output_points} component used in the TPIE raster transformation algorithm.
  }
  \label{fig:generate-output-points}
\end{figure}

\pipefeature{Memory management}
As mentioned, TPIE automatically manages memory and divides available memory among components in a pipeline. Thus in the pipeline definition in Figure~\ref{fig:project-code} there is no code at all dealing with memory distribution. Often many components use only a small amount of static memory, whereas components such as sorting require dynamically allocated memory depending on the amount of available memory. In the latter case the component has to specify its minimum and maximum memory requirements in its implementation. In the example component shown in Figure~\ref{fig:generate-output-points} no requirement is specified since only static memory is used.

\pipefeature{Metadata}
Often many components in a pipeline need some sort of metadata about the items that are being streamed between components. In the example, certain components need to use the dimensions of the input and output rasters. While such metadata can of course be passed as parameters to the individual components in the pipeline definition, doing so makes the definition needlessly cluttered. Instead, TPIE provides a general facility for passing metadata between pipeline components. Thus, in Figure~\ref{fig:project-code} \tplines{forward}, the pipeline definition uses \code{forward()} to pass the dimensions of the input and output rasters to the components that need them. The individual components can then obtain the metadata using \code{fetch()}, such as when the component \code{generate_output_points} in Figure~\ref{fig:generate-output-points} \nodeline{fetch} retrieves the dimensions of the output raster.

\pipefeature{Progress reporting}
In the example, the TPIE support for progress reporting (e.g.\@ as a progress bar) is also used. As with memory requirements, the code required to supply progress information is not part of the code in Figure~\ref{fig:project-code} where the pipeline is defined, but rather part of the implementation of the individual component. Thus, the component  in Figure~\ref{fig:generate-output-points} provides the needed information by using \code{set_steps()} in \nodeline{set-steps} in the \code{propagate()} method to define how many items it will produce, and then calling a progress stepping function in \nodeline{step} in the \code{go()} method for each item that it produces. When executing the pipeline in \tpline{exec} of Figure~\ref{fig:project-code} the argument \code{pi} is a reference to a progress indicator object that tells TPIE how to display progress. To provide accurate progress estimations, TPIE actually uses statistical information about progress of previous runs of the code. To store information about runs, the problem's instance size \code{n}, computed in \tpline{n}, as well as a symbol \code{TPIE_FSI} used to identify the application being executed, are also passed to the TPIE framework when executing the pipeline in \tpline{exec}.

\pipefeature{Parallelism}
The example takes advantage of TPIE support for multi-core CPU parallelism in two ways. First, the sorter components automatically parallelized in the run-formation phase. Second, wrapping the component \code{compute_}\allowbreak\code{trans}\-\code{formation} in the directive \code{tpie::}\allowbreak\code{pipelining::}\allowbreak\code{parallel(...)} in \tpline{parallel} makes TPIE automatically distribute this part of the computation among all available CPU cores.

\section{TPIE Pipelining}
\label{sec:concepts}

In this section we describe the TPIE pipelining framework in more detail. First in Section~\ref{sec:pipelining-use} we describe how to use the framework,
and then in Section~\ref{sec:pipelining-impl} we discuss some aspects of the implementation of the framework.

\subsection{Pipelining Use}
\label{sec:pipelining-use}

In this section we first describe how a TPIE pipeline consisting of a number of components can be modeled using a so-called flow graph, and how this graph can be used to identify pipeline phases. Then we describe how components are implemented. Finally, we describe how a TPIE pipeline is constructed and executed.

\newcommand{\defterm}[1]{\textbf{\emph{#1}}}

\pipefeature{Flow graph and phase identification}
As described in Section~\ref{sec:example-tpie}, a TPIE pipeline consists of a number of components that push data to or pull data from other components. We distinguish between two types of components, namely \emph{regular components} that produce output as the input is processed and \emph{blocking components} that have to process all input before producing any output. Blocking components consist of two sub-components, namely an input and an output sub-component, where the input sub-component processes all input before the output sub-component is invoked to produce the output; the input sub-component might store intermediate results on disk. Sorting is an example of a blocking component; the input sub-component naturally corresponds to the initial run formation step and the output sub-component to the merging step. Blocking components introduce the need for \emph{pipeline phases} that are executed independently, since the input and output sub-components cannot be executed simultaneously. In the raster transformation algorithm in Section~\ref{sec:example}, the three phases are exactly needed due to the two blocking sorting components.

A pipeline can conveniently be represented by a directed acyclic \emph{flow graph}, with \emph{regular nodes} corresponding to regular components and \emph{input nodes} and \emph{output nodes} corresponding to the sub-components of blocking components. Regular nodes are connected with other regular nodes and input and output nodes by edges directed along the streaming direction and labeled as \emph{push edges} or \emph{pull edges} in a natural way; note that a node cannot have both an outgoing push and an outgoing pull edge. Each input node is also connected with a directed \emph{blocking edge} to the corresponding output node.
To be able to automatically determine the phases of a pipeline, TPIE requires that the flow graph corresponding to the pipeline has two particular properties: First, if all blocking edges are removed, then no input and output nodes corresponding to the same blocking component should be in the same connected component. Second, if all push and pull edges are contracted, then the graph should be acyclic. The first property means that the connected components directly identify the pipeline phases that need to be executed independently. When each such connected component is contracted, each directed edge $(u,v)$ in the resulting graph indicates that the phase corresponding to $u$ needs to be executed before the phase corresponding to $v$. Thus the second property ensures that there exists a valid (topological) order in which to execute the components.

While the second flow graph property above has to be
fulfilled for any pipelined program to be valid, the first property only has to be fulfilled if we require that the
program is constructed such that all but blocking components can be
pipelined, that is, that streaming items are only written
to disk by blocking components.
For example, consider the small pipeline shown on the right, where component $u$ pushes to both a sorter and to a component $w$, which in turn pulls output from the sorter. In this case, $u$, $w$ and
\begin{wrapfigure}{r}[0pt]{0pt}\includegraphics{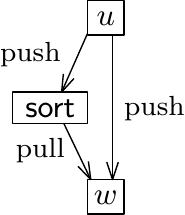}\end{wrapfigure}
the input and output nodes of the sorter are all in the same connected component in the phase graph without blocking edges. However, it is obviously not possible to pipeline all the components in one phase. In particular, it is not possible to pipeline $u$ and $w$, since the output from the sorter used in $w$ is not available at the same time as the output from $u$ also used in $w$. To remedy this problem, and make the flow graph fulfill the second property, a simple blocking component that delays the stream of items from $u$ to $w$, by writing them temporarily to disk, can be inserted between $u$ and $w$ in the pipeline (such that the example has two phases). To support this, TPIE not only contains a built-in sorting blocking component, but also blocking components that \linebreak delay and reverse a stream. Each of these components come in an \emph{active} and a \emph{passive} variant. In the active variant both the input and output sub-components use push-based streaming, and in the passive variant the input sub-component is push-based and the output sub-component pull-based.

\pipefeature{Component implementation}
The interface of each component in a TPIE pipeline must have certain methods; sub-components are essentially like regular components, so when we refer to components below we mean regular components and sub-components.

Methods \code{push()}, \code{pull()}, and \code{can_pull()} are used to stream data items between components. Consider a component corresponding to a node $u$ in the flow graph that produces data that is processed by a component corresponding to a node $v$ in the graph, that is, where there is an edge $(u,v)$. With a slight abuse of notation, we use $u$ and $v$ to refer to the two components. If the edge is a push edge we say that $v$ is a \emph{destination} of $u$, and then $v$ must implement a \code{push()} method and $u$ must push each item in the stream to $v$ by calling the method $v$\code{.push()}. If on the other hand the edge is a pull edge we say that $u$ is a \emph{source} of $v$, and then $u$ must implement a \code{pull()} method and $v$ must pull each item from $u$ by calling the method $u$\code{.pull()}; $u$ must also implement the method \code{can_pull()} to return \code{true} if there is more data to pull and \code{false} otherwise.

A component $u$ that is neither the destination or the source of any other component must implement a \code{go()} method that repeatedly pushes or pulls data until there is no more data to process. This is because data is neither pushed to or pulled from $u$ by other components calling $u$\code{.push()} or $u$\code{.pull()}. Thus the \code{go()} method is used to start the execution of a phase.

Each component $u$ must implement (possibly empty) \code{begin()} and \code{end()} methods that are called before and after the stream processing of the phase containing $u$, respectively. These methods can for example be used to allocate and deallocate memory used by $u$, or set up data structures needed by $u$. Component $u$ is also allowed to push items to its destinations and pull items from its sources in \code{begin()} and \code{end()}. This is e.g.\@ useful when buffers need to be filled up at the beginning or emptied at the end of a phase.

Each component $u$ must also implement a (possibly empty) \code{propagate()} method that is also called before any stream processing in the phase containing $u$ and used to pass metadata between components. Inside the \code{propagate()} method $u$ may use the \code{forward} function to pass key-value pairs to components $v$ that can be reached from $u$ in the flow graph. It may use the \code{fetch} function to retrieve named metadata from other components.
Often metadata includes information about input and output data size, and if TPIE should provide progress reporting, at least one component $u$ in each phase should provide progress information by calling the function \code{set_steps(}$n$\code{)} inside the \code{propagate()} method to indicate the number of items $n$ that it will process, and then call the \code{step()} function once for each item that is processed. Often $u$ is a node with no incoming push or pull edges in the flow graph, that is, the node that creates streaming data.

Finally, a component $u$ that requires dynamically allocated memory to perform its stream processing needs to indicate this to TPIE by calling the functions \code{set_}\allowbreak\code{minimum_}\allowbreak\code{memory(}$a_u$\code{)} and \code{set_}\allowbreak\code{maximum_}\allowbreak\code{memory(}$b_u$\code{)} in its class constructor to request between $a_u$ and $b_u$ bytes of memory, where $b_u = \infty$ is used to indicate that the component requests as much memory as possible. After TPIE has distributed memory, $u$ can then obtain information about how much memory it was assigned between $a_u$ and $b_u$ by calling the function \code{get_available_memory()} in the \code{begin()} method.

\pipefeature{Pipeline construction and execution}
After defining the pipelining components, a pipeline is constructed by stringing together components using the so-called pipe operator as in the expression
  \code{p = generate_output_points() | }\allowbreak\code{compute_transformation() | }\allowbreak\code{sort_S1.input().memory(2)}
where three components, \code{generate_}\allowbreak\code{output_}\allowbreak\code{points()}, \code{compute_}\allowbreak\code{trans}\-\code{for}\-\code{mation()} and the input sub-component of \code{sort_S1} are pipelined. For each component, as for the \code{sort_S1.input()} component in the example, one can set a memory priority using \code{memory()} to indicate to TPIE how important it is to allocate memory to the component; by default the priority is one, and a priority of $k$ means that if several components all request as much memory as possible using \code{set_maximum_memory(}$\infty$\code{)} then a component with priority $k$ will receive $k$ times the amount of memory as one with priority one.

To execute the pipeline one simply calls the object $p$. TPIE then builds the flow graph and computes connected components to identify phases, and then contracts the components and topologically sorts the graph to find the order in which to execute the phases. After this TPIE executes each phase in turn. To execute a phase, TPIE first distributes memory to each component in the phase based on the memory requests and priorities, and then it calls the methods \code{propagate()}, \code{begin()}, \code{go()} and \code{end()} on the components in a specific order based on the flow graph.
First, \code{propagate()} is called on the components in the phase in topological order to allow each component to call \code{forward()}, \code{fetch()} and \code{set_steps()}. The topological order is used since a component $u$ has to be able to pass metadata to components reachable from $u$ in the flow graph.
Next \code{begin()} is called on all components in the order obtained by topologically sorting the flow graph where all push edges have been reversed; this topological order exists as the graph is acyclic, since a node in the flow graph cannot have both an outgoing push and an outgoing pull edge. This particular topological order is used since for a push edge $(u,v)$, $u$ should be able to push to $v$ in $u$\code{.begin()}, so $u$\code{.begin()} should be called after $v$\code{.begin()}; similarly, if $(u, v)$ is a pull edge, then $v$\code{.begin()} must be called after $u$\code{.begin()} is called.
After this initialization, the main streaming part of the phase is executed by calling the \code{go()} method on the appropriate component.
Finally, at the end of the phase \code{end()} is called on the components in reverse order of the \code{begin()} order, that is, in reverse topological order.

\subsection{Pipelining Implementation}
\label{sec:pipelining-impl}

Above we have already discussed how TPIE identifies and executes phases of a pipeline, and due to space constraints we cannot describe the entire implementation of TPIE pipelining in detail. In this section we therefore briefly discuss a few aspects of the implementation not described above. The interested reader is also referred to the technical documentation of TPIE
.

\begin{figure}[t]
  \centering
\newcommand{\memorytablelambda}{2}
\newcommand{\memorytableassigned}{36}
\def\memorytablerows{%
$A$  & $4$            & $12$           & $\textbf{5}$   & $10$ \\
$B$  & $1$            & $7$            & $\textbf{3}$   & $6$ \\
$C$  & $\textbf{8}$   & $\infty$       & $3$            & $8$ \\
$D$  & $7$            & $\textbf{12}$  & $7$            & $12$ \\
}
  \begin{tabular}[b]{ccccc}
    \hline
    Node & Minimum      & Maximum       & Priority     & Assigned \\
    $v$  & $a_v$        & $b_v$         & $c_v$        & $M_v(\lambda)$ \\
    \hline
    \memorytablerows
    \hline
  \end{tabular}
  \par
  \vspace{1ex}
  \includegraphics{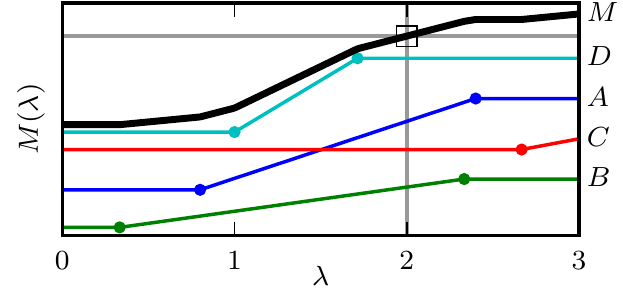}

  \caption{
    Memory assignment for four nodes where $\lambda = \memorytablelambda$,
    $M(\lambda) = \memorytableassigned$.
    Nodes $A$ and $B$ are assigned $\lambda$ times their priority,
    whereas nodes $C$ and $D$ are assigned their minimum and maximum memory,
    respectively.
  }
  \label{fig:memory}
\end{figure}

\pipefeature{Memory management}
As mentioned, TPIE distributes memory to components in a phase based on the minimum $a_u$ and maximum $b_u$ memory requirements, along with the memory priority $c_u$ of each component $u$ in the phase. Component $u$ is assigned $M_u(\lambda) = \max\{a_u, \min\{b_u, \lambda c_u\}\}$ bytes of memory, for a value of $\lambda$ such that the total assigned memory $M(\lambda) = \sum_u M_u(\lambda)$ is smaller than the available memory. This way memory is distributed proportionally to the memory priorities unless this gives an amount of memory outside the $[a_u,b_u]$ interval. Since $M(\lambda)$ is a non-decreasing function of $\lambda$, TPIE can use binary search to find $\lambda$.
Refer to Figure~\ref{fig:memory} for an example.

\pipefeature{Progress reporting}
TPIE also supports progress reporting for a pipeline. To ensure that a progress bar shown to the user progresses from 0\% to 100\% at a constant pace, TPIE maintains an \emph{execution time database} with information about how large a fraction of the execution time was spent in each phase in the execution of the pipeline that processed the largest instance size. In order to distinguish between different pipelines in the execution time database, each pipeline execution carries the preprocessor macro \code{TPIE_FSI} as an argument, which is expanded by the compiler into a string that uniquely identifies the location in the code where the pipeline is defined. In this way, TPIE can store information about multiple pipelines in the execution time database.

\pipefeature{Automatic parallelization}
Automatic multi-core CPU parallelism is supported in TPIE by applying the \code{tpie::pipelining::parallel(...)} directive to a push-based pipeline component. In this case, the processing of the component is distributed among the available CPU cores by instantiating a copy of the component for each core and passing items to these components in a round-robin fashion as they arrive.
To amortize the overhead of thread synchronization, items are passed to components in batches of $2048$ items at a time.

\pipefeature{Function inlining}
To minimize the computational overhead of the many \code{push()} and \code{pull()} function calls required when executing a pipeline, TPIE is designed to allow the compiler to inline the processing of several consecutive pipelining components into one function. As in STXXL, this is achieved using \emph{template instantiations} in {\Cpp}. Thus, if $u$ pushes to $v$ in the pipeline, then when $u$ is compiled the type of $v$ is known to the compiler so the implementation of $v$\code{.push()} can be inlined into $u$. However unlike STXXL, TPIE is designed to hide the resulting complex type definitions from the pipeline definition, and instead of using recursive template instantiations the pipe operator can be used to define a TPIE pipeline. This is very convenient when building large pipelined applications.

\clearpage
\end{document}